# The Normal, the Natural, and the Harmonic

Theodore Modis[*]

**Abstract**

Use is made of rigorous definitions for the terms normal, natural, and harmonic to reveal a number of unfamiliar aspects about them. The Gaussian distribution is not sufficient to determine who is normal, and fluctuations above or below a natural-growth curve may or may not be natural. A recipe for harmonically sustained natural growth requires that the overlap during the substitution process must be limited. As a consequence the overall growth process must experience good as well as bad "seasons".

*Keywords:* normal; Gaussian; natural growth; fluctuations; logistic growth; harmonic; product substitution; cascading logistics; business seasons.

[*] Theodore Modis is the founder of Growth Dynamics, an organization specializing in strategic forecasting and management consulting. http://www.growth-dynamics.com
Address reprint requests to: Theodore Modis, Via Selva 8, 6900 Massagno, Lugano, Switzerland. E-mail: tmodis@compuserve.com.

## 1. Introduction

The hard sciences have traditionally attached rigorous definitions to common words. Most people are familiar with the physics terms: energy, force, momentum, and power. But more subtle concepts such as action, impulse, natural, and harmonic have also been endowed with scientific rigor. When everyday notions enter the world of science they become subject to natural laws and may occasionally yield unexpected insights.

## 2. To Be Normal Is to Be Different

Consider, for example, the word "normal". To define normal behavior psychologists use the bell-shaped distribution curve named Gaussian after its creator Karl Friederich Gauss (1777-1855). In textbook literature the Gaussian has been traditionally used by hard and soft sciences alike to describe the data sets derived from many applications. Examples range from people's intelligence (I.Q. index), height, weight, and so forth, to the velocities of molecules in a gas and the ratio of red to black outcomes at casino roulette wheels. A person is considered to be normal in a certain aspect as long as he or she is within one standard deviation from the average in the Gaussian distribution of the variable that describes the aspect.

To be within one standard deviation from the average corresponds to a probability of 68.3%. However to be within one standard deviation in two variables, say I.Q. and height at the same time, has a probability of only 46.6% assuming the two variables are independent. To be within one standard deviation in three variables simultaneously, the probability is only 31.9%. The probability goes down rapidly as we add more variables: wealth, looks, health, sports performances, and so on. To be normal in ten different variables, assuming always independence between them, the probability is 2.2%, and there is only one chance in a million that someone is within one standard deviation in 36 different variables, which makes him or her a real freak worthy of a place in a circus. Finally, there should be no one on earth today who is at once normal in 55 or more different independent aspects.

In this light, we may want to revise the definition of a normal person. A person should be normal when he or she is close to the average in only a few aspects but may be far from the average in most other aspects. To be more precise, the probability distribution goes down exponentially with the number of variables. The average of this probability distribution is 3.2 variables and its standard deviation is 0.7. Therefore, to be within one standard deviation here it suffices to be "normal" in only two to four variables. For example, if you are of average intelligence and height, you are normal. But if in addition you are of average, health, wealth, and will power, then you are not quite normal.

By and large it is being different from others that renders one normal.[1]

## 3. Growth in Competition

Gauss's distribution has been extensively used in the sciences always referred to as the normal distribution. However, all this glory to Mr. Gauss is somewhat circumstantial and undeserved. A well-known man of science, he was referred to as "the prince of mathematics" in nineteenth-century literature, and the bell-shaped curve he provided resembled very much the distributions with which people were concerned. In addition, it is obtained through an elegant mathematical derivation. For historical reasons mostly, people have proceeded to put Gaussian labels on most of the variables on which they could get their measuring sticks. However, in reality there are not many phenomena that obey a Gaussian distribution *precisely*. Moreover, there is no natural law behind it.

There is another bell-shaped distribution, very similar to Gauss's curve, which has a better reason for existence. It is the life cycle of natural growth, otherwise known as logistic growth.

Logistic growth describes how a species population grows to fill its ecological niche under conditions of natural competition (survival of the fittest). This fundamental natural law also describes how we learn, and how rumors and epidemic diseases spread. One can imagine Pierre Verhulst (1804 - 1849) reasoning like a physicist when he cast his observations on the growth of species populations into the following equation:

$$\frac{dX}{dt} = aX(M - X) \qquad \text{where } a \text{ and } M \text{ constants} \qquad (1)$$

Given that rabbits procreate, the rate of growth of a rabbit population must be at all times proportional to the size of the population. At the same, given that a grass range can only feed a maximum number of rabbits $M$, the rate of growth of the population must at all times be also proportional to the remaining capacity for rabbit growth $(M - X)$, and progressively go to zero as $X$ approaches $M$.

It is unlikely that Gauss reasoned in a similar manner when he derived his equation for a bell-shaped distribution. Unlike physicists, mathematician purists are not necessarily concerned with natural laws and the effects they have on life.

**The Normal Distribution Is Very Similar to the Natural Life Cycle**

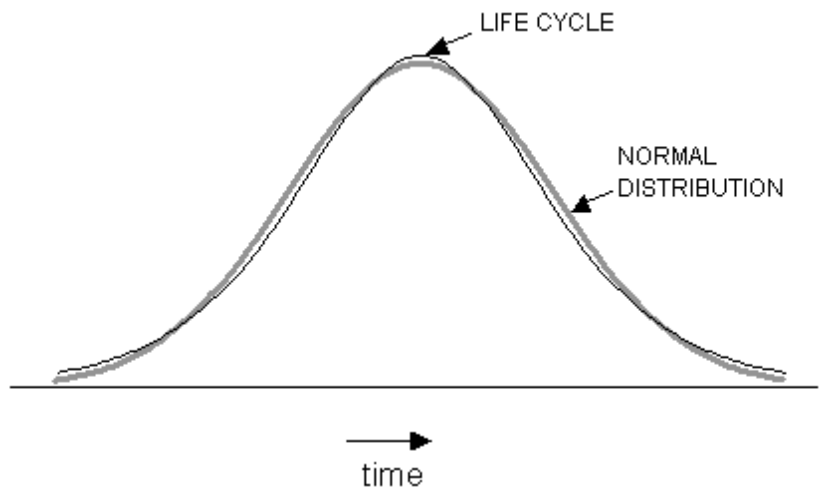

Figure 1. Comparison of the Gaussian — normal distribution — and a natural life cycle. The area under the two curves is the same.

## 4. Natural and Unnatural Fluctuations

The solution of Equation (1) is the familiar S-shaped logistic curve, often referred to as the natural-growth curve that has become the pattern of all growth in competition.

But when Equation (1) is put in the form of a difference equation, it becomes

$$X_{n+1} = r X_n (1 - X_n) \qquad \text{where } r \text{ is a constant} \qquad (2)$$

This equation is strikingly similar to Equation (1), but whereas Equation (1) is solved via integration and its solution gives rise to the smooth S-shaped logistic pattern, Equation (2) is solved via iteration and its solution gives rise to the same pattern for low values of $r$ but may also give rise to states of chaos for higher values of $r$ ($3.772 < r < 4.0$). The former emphasizes the presence of a trend and has become the tool to describe natural growth. The latter emphasizes the lack of trend and has become the tool to describe chaos. The chaotic fluctuations appear on what corresponds to the ceiling of the logistic after the upward trend has died down.

It has also been shown that if Equation (1) is first solved and then made discrete, chaotic-type fluctuations can be expected before as well as after the curve's steep rise.[2] One may in fact argue that in real life, fluctuations in data may exist all along the logistic curve but they may not be visible during the curve's steep rise because they are masked by the pronounced upward trend. If you ask someone with trembling hands to draw an S-shaped curve, you will obtain something like that of Figure 2 where indeed the fluctuations are more visible before and after the steep rise even if the person's hand trembled in a similar way throughout the drawing process. There is, however, a significant difference between fluctuations at the beginning and the end, and fluctuations during the steeply rising part of the logistic. The latter generally correspond to reality whereas the former will never become realized.

**Drawing an S-Curve with Trembling Hands**

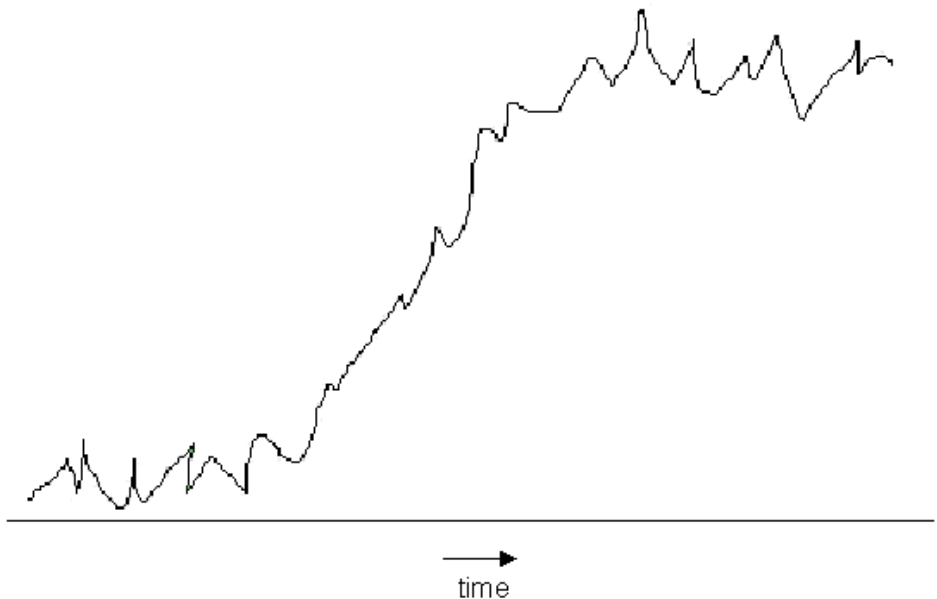

Figure 2. The fluctuations during the steeply rising section are less visible because they are masked by the pronounced upward trend.

Consider a fluctuation above the logistic trend at time $t_1$ during the curve's steep rise as shown in Figure 3. The extremity of this fluctuation corresponds to the curve's level at $t_2$. The only "abnormality" introduced by the fluctuation is that of a precocious appearance, because the level reached is realistic as it will appear in a short time in the future. But the same fluctuation at time $t_3$ reaches a level that will never be achieved by the natural-growth process, and thus has zero chance of being realized. Such a fluctuation can be considered as unnatural. It would be simply unusual if it snowed in Greece in October, but it would seem unnatural if it ever snowed in the Sahara. During irregularities events occur earlier or later than their natural time, but during chaos events have no corresponding time in which they could seem natural.

Whenever the fluctuation reaches a level that corresponds to a time within the gray pattern of Figure 3, the fluctuation can be considered as natural. If the fluctuation's level has no correspondence on the logistic curve, it deserves to be labeled as unnatural fluctuation. At time $t_3$ the fluctuation should have much smaller amplitude to qualify for naturalness unless it was in the downward direction. For a fluctuation to be *natural* its extremities must correspond to points on the curve that map to the gray pattern in Figure 3, which is the natural-growth life cycle — what Gauss approximated with his mathematical concoction.

In this way the *normal* distribution plays a central role in the definition of a *natural* fluctuation.

But this definition, when applied strictly, is not entirely adequate. Naturalness is intuitively associated with the probability of being realized. A downward fluctuation at the ceiling would appear less and less natural the further it occurred from the logistic rise. Even if the level of its extremity had been realized sometime in the past, if this was very long time ago, the fluctuation would seem rather unnatural. And this is the reason that chaotic fluctuations are to be treated as generally unnatural. They are superimposed on the level of the logistic's ceiling and they are generally far removed from the initial rise to that level. Whether they are upward or downward they appear unnatural because a realization of this value via a natural-growth process is either impossible or it took place a very long time ago.

## Two Types of Fluctuations

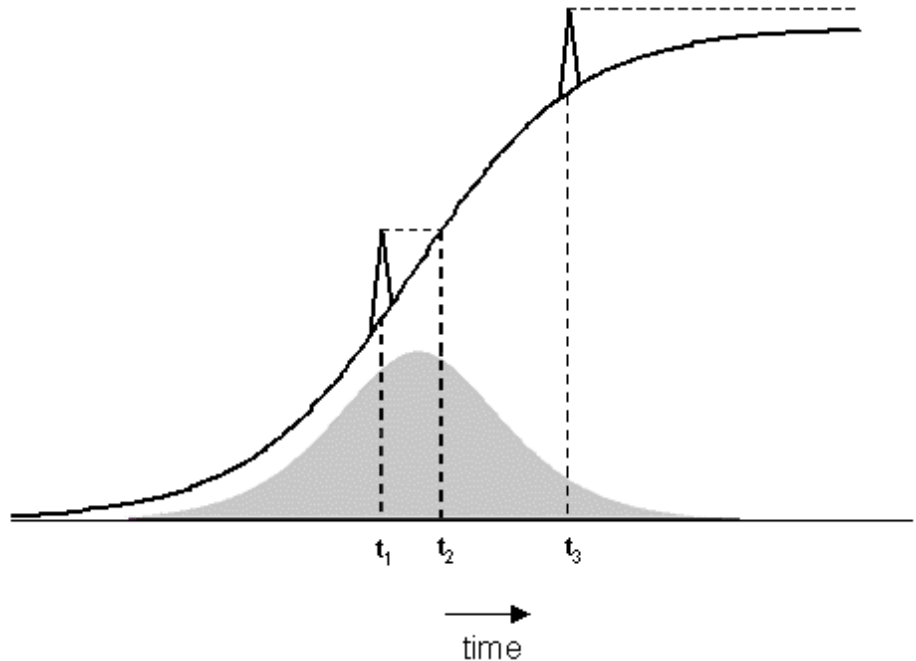

Figure 3. A fluctuation at $t_1$ reaches the same level as the logistic curve at $t_2$. The same-size fluctuation at time $t_3$ has no corresponding point on the curve. The gray life cycle delimits the position and size of all *natural* fluctuations. If the fluctuation at $t_3$ was much smaller, it could have been natural.

### 5. Deciding on the Time of the Replacement

Marketers have long used bell-shaped curves to describe the life cycles of their products. After all, product sales do grow in a competitive environment and the best-fit product normally wins.

But marketers have a harder time accepting the fact that natural growth is capped and that their company will not continue growing exponentially, not even linearly. The only way growth can be sustained in a competitive environment is via a succession of S-shaped logistic steps as one market niche is filled after the other. Even then, the overall envelope will follow a logistic pattern on a larger scale.[3]

A well-known marketers' utopian pursuit is the strategy of product replacement timed so as to avoid any slowdown in the growth of their sales revenue. Evidently, launching products too closely together (the case may be argued for new Microsoft Windows operating systems) may frustrate customers and/or lead to "cannibalization" of their own market when the new product robs sales from the old one. On the other hand, delaying the launching of a replacement product may create a vacuum in a vendor's offerings and result in loss of customers to the competition. So the question becomes when is the *optimum* time to launch a replacement? The question can be generalized to: when is the right time to introduce change in an ongoing natural-growth process? No one wants to tamper with something that works well, but how old should become a product before its replacement is launched?

One criterion can be found in harmonic motion and not only because the concept of harmony implies goodness. Regularly spaced product life cycles produce a landscape pattern suggestive of a sine wave, and large-scale growth processes, such as world energy consumption, have deviated from a natural trend so as to also produce a sinusoidal wave (Kondratieff's cycle).[4] The sinusoidal pattern is characteristic of the pendulum's harmonic motion.

The cascade of two identical logistics was studied as a function of the distance in time between them for different time constants. The logistics, representing cumulative sales, were cascaded in a mutually exclusive way, that is, the first logistic stopped when the second took over. This was essential in order to maintain an approximately constant market in terms of average sales per unit of time. Otherwise, we would not be addressing the replacement of products but the expansion markets via rapid product launching, which is a different and rather debatable subject.

**A Harmonic Cascade of S-curves**

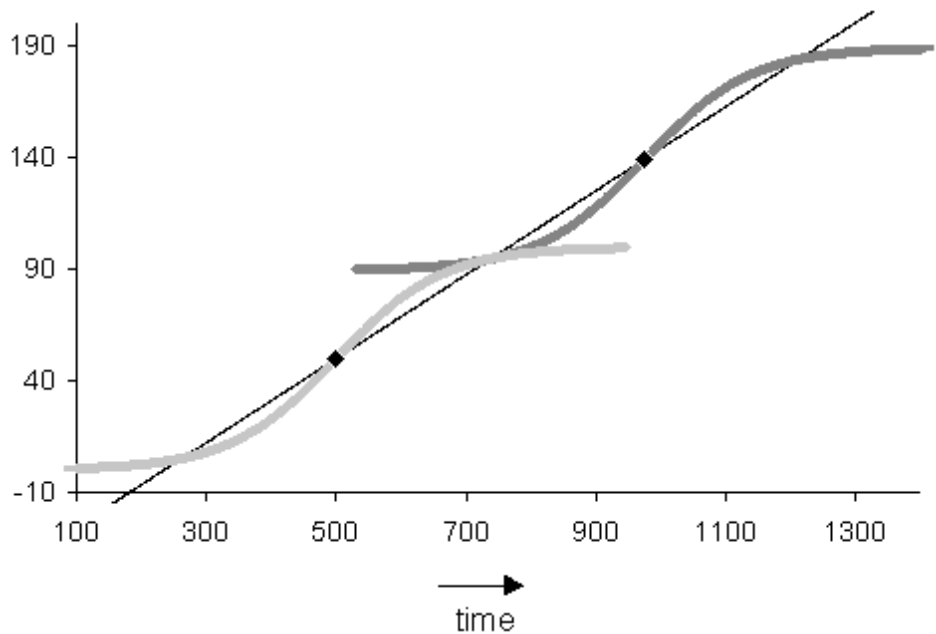

Figure 4. The same logistic pattern repeats (dark gray line) from where the previous one (light gray line) leaves off. The straight line connects the center points of the two logistics. The axes have arbitrary units.

In Figure 4, a straight line connects the centers of the two logistics and thus serves as a general trend. The difference between the envelope of the cumulative sales and the straight line was extracted and fitted by a sinusoidal wave. The fit involved four parameters: the time delay between the two logistics, and the amplitude, frequency and phase of the sine wave. The fitting process minimized the sum of differences squared, akin to a Chi Square. The fit was excellent as can be appreciated in Figure 5. It was also stable against a factor-of-ten change in the time constant of the logistic used.

There are two reasons for which it was decided to fit the deviation from a straight-line trend in the cumulative-sales pattern rather than to fit the sum of the two sales life cycles.

One reason is that the outline of the two life cycles, despite the fact that it depicts peaks and troughs, it is not quite sinusoidal as it comes to a point when the two curves meet (see Figure 6); it would not result in a good fit. A second reason is that the fitting procedure would require five parameters to be varied instead of four, as a flat variable level would have to be added to the sine wave.

The decision to fit the deviation from a trend rather than the life cycles led to a displacement of the peaks. The two life-cycle peaks occur at times 500 and 976 (Figure 6), whereas the two corresponding overall-envelope maxima are at times 618 and 1093 respectively (Figure 5). Such a time lag is corroborated by an observation on Kondratieff's cycle based on energy consumption cited in Reference [4], where the cycle's maxima appear almost ten years later than the corresponding peaks in the rates of growth.

**Cascaded Logistics Match A Harmonic Wave**

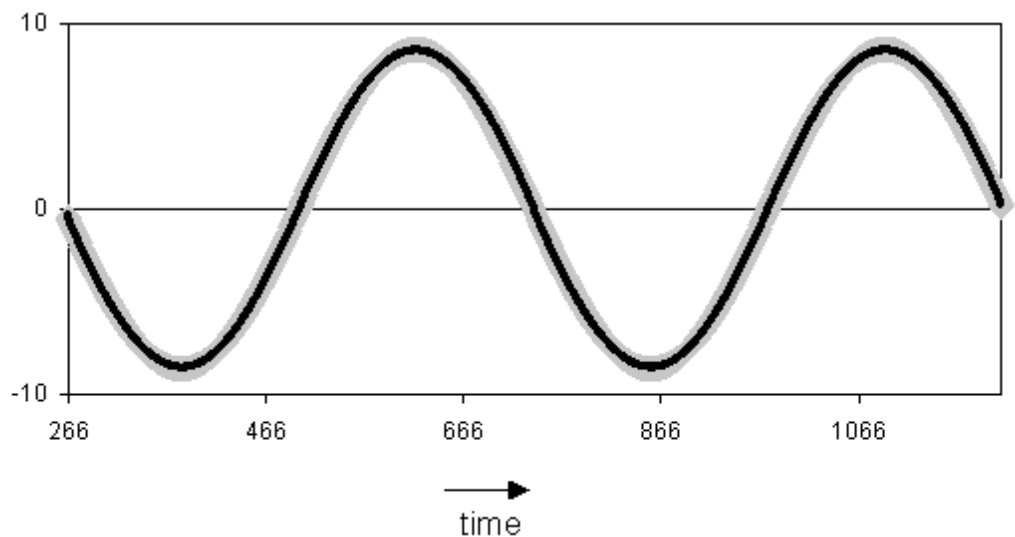

Figure 5. The difference between the logistics envelope and the straight line in Figure 4 is shown here in black. The thicker gray line is a fitted harmonic wave (sinusoidal). The axes show the same arbitrary units as in Figure 4. The crossover from one logistic to the next takes place at time = 738 and shows no perceptible deviation from the smooth harmonic wave.

The overlap of the cascading logistics is rather small: 2.6%, 10.3%, and 32.7% of the replacement logistic coincide with 89.7%, 97.4, and 99.4 respectively of the incumbent logistic, see Figure 6.

This recipe spells out in a quantitative way how to *harmonically* cascaded *natural*-growth processes. It can be used to time the launching of new products using only the sales information at hand. In order for the new product to achieve 2.6% penetration of its market when the old product reaches 89.7% penetration of its market, the launching of the new product — corresponding to the nominal beginning of the new logistic, namely its 1% level — must occur halfway down the phasing-out period of the incumbent product. This point in time falls between the two little diamonds on the declining side of the life cycle in Figure 6.

Like farmers sowing in the fall the seeds for the next crop, product managers must sow the seed of their new product during the "fall season" of the incumbent product.

## 6. If Winter Is Here, Can Spring Be Far Behind?

Many management theorists divide the growth cycle — typically a product's sales cycle — into segments. Theorists generally consider four periods according to the phase of growth: start-up, rapid growth, maturation, and decline. Their treatment is invariably qualitative, and the four phases are not necessarily of equal duration or precise definition.

But here the business cycle is presented somewhat differently, using the four seasons as a metaphor. Winter reflects the critical growth period encountered during the beginning and the end of a natural growth process. Products experience two winters in their lifetime. The first winter is while they are struggling for a foothold in the marketplace, and the second one when they are exiting the market and the follow-up product is fighting for succession.

The end of the first winter — that is, when the product has realized around 10 percent of its growth potential — signals that the growth process has survived "infant mortality". Summer season appears around the peak of the rate of sales, whereas a second winter season appears around the time of product replacement. In-between there is progressively rising and progressively declining rates of sales corresponding to spring and fall seasons respectively. In Figure 6 we see markings (little diamonds) showing the limits of business seasons defined in this way. The four seasons have the same duration that is equal to ¼ the time delay between the two logistics.

A seasons metaphor has more than poetic justification (this section's title is from a poem by Percy B. Shelley). Our familiarity with the mechanisms associated with nature's four seasons can shed light on and guide us through decisions on business and social issues. For example, the low creativity observed during summer is only partially due to the heat. New undertakings are disfavored mainly because summer living is easy and there is no reason to look for change. In contrast, animals (for example, foxes and sparrows) are known to become entrepreneurial in winter.

Often what naturally happens is what should happen. As strange as it may sound, seeing specialists progressively evolve into bureaucrats may be a good sign. It is one indication that summer is setting in. The word *bureaucrats* carries a negative connotation, but if we call them *process agents* instead, we realize that they provide an important function during times of high growth and prosperity. It is during summer that enterprises become successful, centralized, conservative (no one tampers with something that works well), and in need of clockwork operations. Fine-tuning and zero defects (the original aspiration of total quality management) are particularly appropriate for a summer season.

In contrast, excellence is not a top priority during winter; fundamental change and business process reengineering are at the top of the list. Winter is the most difficult but also the most fertile season. Despite low morale, innovation and creativity are at a high. New directions are set. Mutations come out in great numbers and compete for the next position in power. Most of them will die, but those that make it to springtime will be ensured of a full growth cycle. Mutations serve the purpose of emergency reserves. In industry they can take the form of new product ideas, basic innovations, or other ventures. The higher their number, the better the chance that some of them will survive and grow, paving the way for the sunnier seasons that lie ahead. Winter is a period of selection during which only the best-fit mutations will survive. The best-suited human resource for the winter is the entrepreneur.

In the winter one may raise fundamental questions such as *why* embark on some endeavor, whereas in the spring the questions are around *what* to do (the product), and in the fall on *how* are things done (the processes). There are many attributes and appropriate behaviors for each season. They have been studied extensively and applied to situations that depict succession of seasonlike stages ranging from corporate profitability and product sales to the introduction of change in personal affairs.[5]

What we want to retain here is that if natural-growth processes are cascaded in a *harmonic* way as in Figures 4-6, then there will be a periodic and regular swing between good and bad business seasons. This swing is not simply natural and inevitable. It is also desirable because it will play a significant role in triggering new growth, just as pruning the roses ensures healthier blossoms for the next season. The lesson for business executives facing a major transition between products, technologies, or other fundamental change is not to strive for minimizing its impact but to plan for and anticipate a low-growth period comparable in duration to the high-growth period the just enjoyed.

**The Overlap between the Two Logistics of Figure 4**

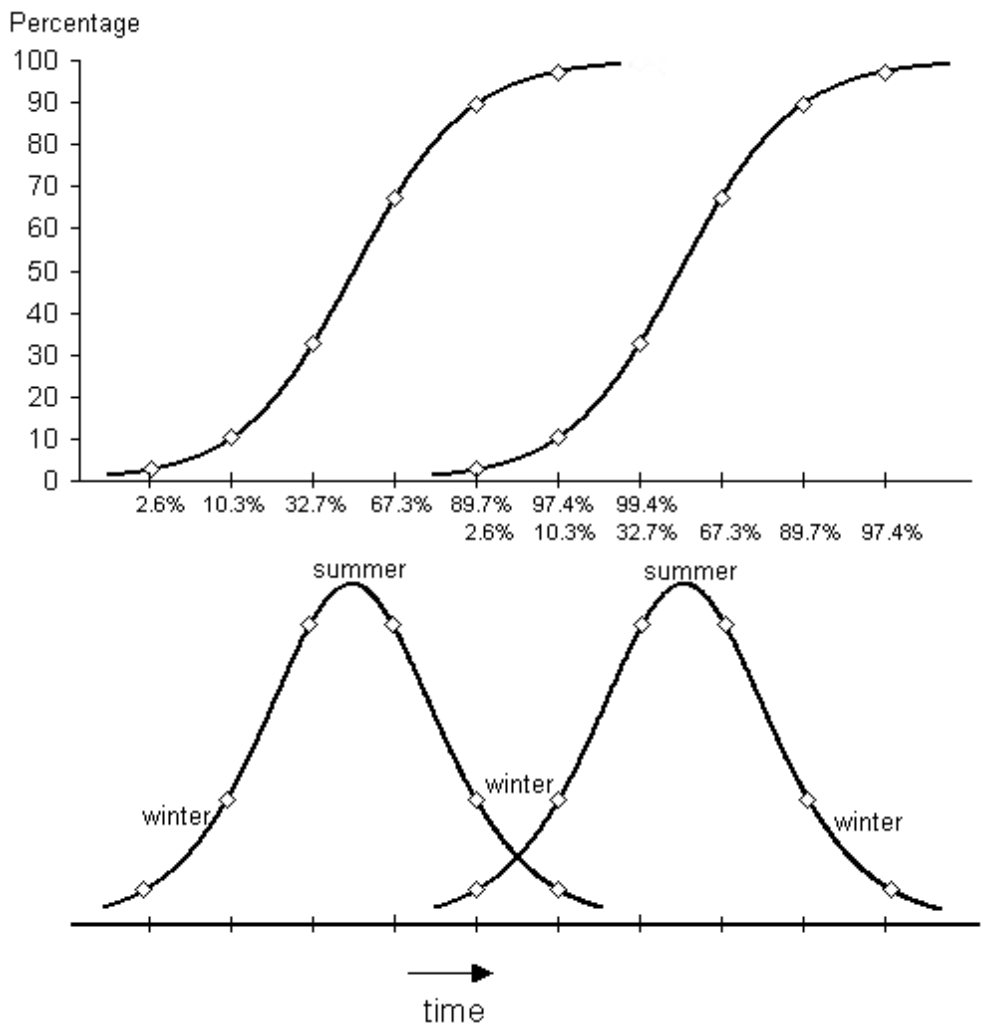

Figure 6. The two logistics of Figure 4 and their corresponding life cycles. The horizontal axis at the middle shows percentage penetration levels for each logistic. The horizontal axis at the bottom shows the same arbitrary time units as in Figures 4 and 5. The little diamonds delimit business seasons concerning individual-product sales, see text.

Proverbial wisdom has long claimed that there is goodness in every season. Some people may think that the most desirable climate is found on tropical islands like Mauritius and Seychelles. Not true! Our above argument based on harmony dictates a large and regular seasonal variation like that encountered in temperate climate, which has also been the cradle of most great civilizations. History is poor in significant cultures that emerged from the arctic or from the tropics, the former perhaps because of conditions hostile to life and the latter mostly because of lack of variation and motivation. Despite an idyllic setting, tropical islands are rather sterile.

The expulsion of Adam and Eve from Paradise may have been, after all, the original blessing.